\documentclass[twocolumn,conference]{IEEEtran}
\usepackage[LGR,T1]{fontenc}
\usepackage{float}
\usepackage{amsmath}
\usepackage{amssymb}
\usepackage{graphicx}
\usepackage[unicode=true,
 bookmarks=true,bookmarksnumbered=true,bookmarksopen=true,bookmarksopenlevel=1,
 breaklinks=false,pdfborder={0 0 0},pdfborderstyle={},backref=false,colorlinks=false]
 {hyperref}
\hypersetup{pdftitle={Your Title},
 pdfauthor={Your Name},
 pdfpagelayout=OneColumn, pdfnewwindow=true, pdfstartview=XYZ, plainpages=false}
\usepackage{breakurl}

\makeatletter

\DeclareRobustCommand{\greektext}{%
  \fontencoding{LGR}\selectfont\def\encodingdefault{LGR}}
\DeclareRobustCommand{\textgreek}[1]{\leavevmode{\greektext #1}}
\ProvideTextCommand{\~}{LGR}[1]{\char126#1}

\newcommand{\lyxmathsym}[1]{\ifmmode\begingroup\def\b@ld{bold}
  \text{\ifx\math@version\b@ld\bfseries\fi#1}\endgroup\else#1\fi}

\providecommand{\tabularnewline}{\\}

\usepackage[caption=false,font=footnotesize]{subfig}

\makeatother

\begin{document}

\title{Multilateration of the Local Position Measurement}

\author{\IEEEauthorblockN{Juri~Sidorenko, Norbert~Scherer-Negenborn, Michael~Arens, Eckart~Michaelsen}\IEEEauthorblockA{Fraunhofer Institute of Optronics, System Technologies and Image Exploitation
IOSB\\
Gutleuthausstrasse 1, 76275 Ettlingen, Germany. \\
Juri.Sidorenko@iosb.fraunhofer.de}}
\maketitle
\begin{abstract}
The Local Position Measurement system (LPM) is one of the most precise
systems for 3D position estimation. It is able to operate in- and
outdoor and updates at a rate up to 1000 measurements per second.
Previous scientific publications focused on the time of arrival equation
(TOA) provided by the LPM and filtering after the numerical position
estimation. This paper investigates the advantages of the TOA over
the time difference of arrival equation transformation (TDOA) and
the signal smoothing prior to its fitting. The LPM was designed under
the general assumption that the position of the base station and position
of the reference station are known. The information resulting from
this research can prove vital for the system\textquoteright s self-calibration,
providing data aiding in locating the relative position of the base
station without prior knowledge of the transponder and reference station
positions.

\end{abstract}

\IEEEpeerreviewmaketitle{}

\section{Introduction}

In the past, a wide range of different methods and sensors have been
developed to obtain the exact position of an object of interest. The
most common are radio frequency based methods, like NAVSAT GPS. This
technology is often used as an example for the time of arrival equation
(TOA). The traveling time between the satellite and the sensor on
the ground can be used to estimate the sensor\textquoteright s position,
neglecting the position estimation by phase. Both, the satellite\textquoteright s
position in its orbit and the signal\textquoteright s send time, are
known. Combining this information with the time signal arrival on
the ground and the speed of light, it is then possible to estimate
the range between the satellite and the sensor. This range is called
pseudo range. If we neglect the time offset, three satellites are
required to estimate the sensor\textquoteright s 3D position.\\
Unfortunately, the data update is quite slow and unsuitable for urban
territory. During WWII, TDOA systems like DECCA became very popular.
The TDOA method does not require knowledge of emission times. In contrast
to the TOA, all possible locations for one measurement are located
on a hyperbola. Other methods, like the angle of arrival, will not
be addressed in this paper. All examples used from here on out will
be based on the Abatec LPM. This system has potential, due to the
fact that it is able to operate both in- and outdoors and provide
an update rate of 1000 measurements per second.\\
\begin{figure}[H]
\begin{centering}
\includegraphics[scale=0.28]{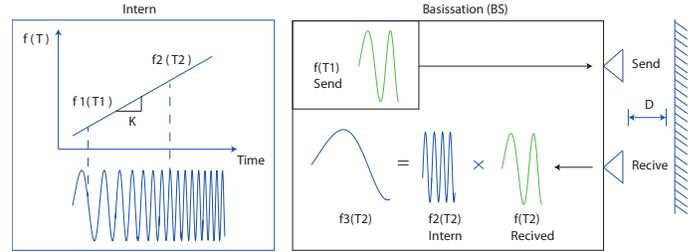}
\par\end{centering}
\caption{FMCW\label{fig:FMCW}: Left increasing frequency chirp with slope
K. Right: mixed frequencies }
\end{figure}
 
\begin{figure}[H]
\begin{centering}
\includegraphics[scale=0.28]{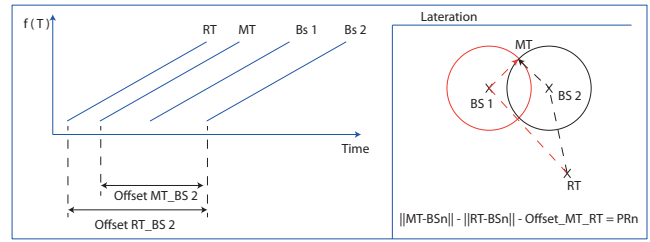}
\par\end{centering}
\caption{LPM\label{fig:LPM} Left: Offset difference between transponder MT,
basis station BS and reference station RT. Right: LPM euclidian equation}
\end{figure}
The LPM was highly inspired by the FMCW frequency modulated continues
wave radar (FMCW) and systems based on TOA, such as GPS. Figure \ref{fig:FMCW}
illustrates the functionality of a FMCW. The internal system is generating
an increasing frequency chirp with a fixed slope K. This chirp is
sent and reflected by an object. The received signal does not change
its frequency, but during this time $(T2-T1)$ the internal frequency
changed from $f1$ to $f2$. The radar can use an additive or multiplicative
mixer to obtain the sum of the two frequencies and their differences.
As only the frequency difference is required at this point, the frequency
sum can be filtered by a low pass filter. The LPM is using this principle
already, but in contrast to the FMCW radar, the send frequency chirp
is getting compared with the frequency chirp of the other base station
$KBS1=KBS2$ (figure \ref{fig:LPM}). If the chirps are synchronized
(started at the same time) there is no offset, but since the base
stations and transponder are not synchronized, a reference station
is required. Every frequency measurement the LPM provides is based
on the difference between the transponder and one of the base stations
with respect to the reference station at the same base station. Due
to this fact, the time offset O is equal for every base station for
the same measurement as demonstrated in fig. \ref{fig:LPM}. This
all leads to the LPM equation (\ref{eq:LPMeq}), with R: Pseudo Range,
O: Offset, M: Transponder- and T: Reference station for different
measurements i.\\
\begin{equation}
R_{i}=O+\left\Vert M-B_{i}\right\Vert -\left\Vert T-B_{i}\right\Vert \label{eq:LPMeq}
\end{equation}

\begin{equation}
\left\Vert M-B_{i}\right\Vert =\sqrt{\left(x_{i}-x_{M}\right)^{2}+\left(y_{i}-y_{M}\right)^{2}}
\end{equation}

\begin{equation}
\left\Vert T-B_{i}\right\Vert =\sqrt{\left(x_{i}-x_{T}\right)^{2}+\left(y_{i}-y_{T}\right)^{2}}
\end{equation}

\section{Previous Work}

The measurement principal of the Abatec LPM and its hardware implementation
have been presented in \cite{Pourvoyeur2005} \cite{Resch2012}. Previously
published work about the LPM predominantly focused on the usage of
Kalman filters to detect an outlier\cite{R.Pfeil2009} or to track
the position of the transponder \cite{Pourvoyeur2006a}. An approach
to outlier detection can be found in \cite{R.Pfeil2009}, were the
linear Kalman filter in combination with the $\chi^{2}$ test was
used to detect outliers within the offset corrupted data. Nonlinear
equation solving with Bancroft is analyzed in \cite{Pourvoyeur2006}\cite{Stelzer2004}
and compared with the least median of squares (LMS) in \cite{Pfeil2009}. 

\subsection*{Abatec LPM (state-of-the-art):}
\begin{itemize}
\item Solving of nonlinear TOA equation with a numerical solver.
\item Unknown variables are coordinates of the transponder and the offset.
\item Filtering after nonlinear multilateration.
\end{itemize}

\subsection*{New approach:}
\begin{itemize}
\item TOA to TDOA transformation.
\item The TDOA data is filtered before the multilateration.
\item Both a linear TDOA solution and a nonlinear TDOA solution may be used
to filter data before the lateration. 
\end{itemize}

\subsection*{Advantages of new approach:}
\begin{itemize}
\item Filtering before solving has the advantage that Gaussian noise inside
of measurement data does not change due to numerical solver.
\item The linear solution is faster than the nonlinear solution. 
\item The nonlinear solution does not have to fit the offset. 
\item Without the offset, outlier detection is enabled.
\end{itemize}

\subsection*{Disadvantage of new approach:}
\begin{itemize}
\item The TDOA solution requires an additional base station 
\item The linear solution is more easily affected by unfavorable conditions
. 
\end{itemize}

\section{Methodology}

\begin{figure}[H]
\begin{centering}
\includegraphics[scale=0.6]{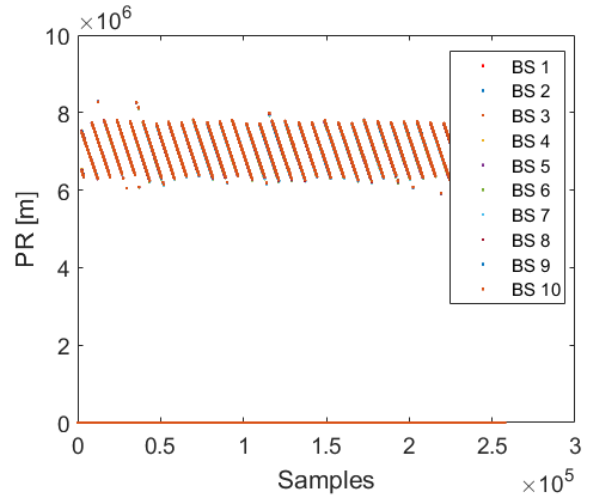}
\par\end{centering}
\caption{\label{fig:LPM-raw-data}LPM real raw data with offset Z at different
measurements}
\end{figure}

\begin{figure}[H]
\begin{centering}
\includegraphics[scale=0.6]{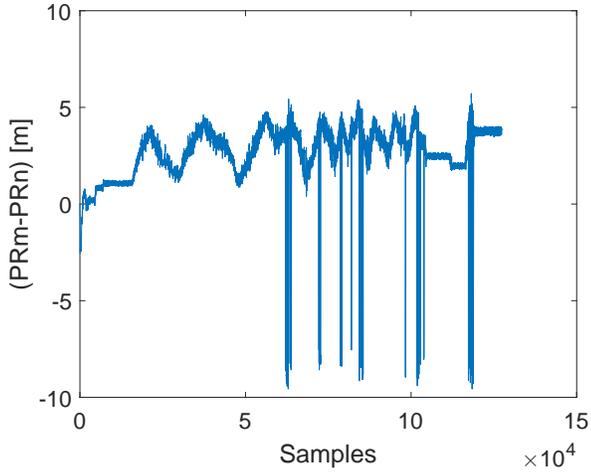}
\par\end{centering}
\caption{\label{fig:Relative-position}Pseudo range of the real data after
the transformation}
\end{figure}

The previous section introduced the LPM and the following focuses
on its position estimation. In general, the exact positions of the
reference station and base stations are known. In this way, four base
stations are required to obtain the x, y, and z coordinates of the
transponder. The euclidean form for this equation can be solved using
either the Gauss-Newton method, or another non-linear solver. Alternatively,
the equation may be linearized with a Taylor-series expansion \cite{FOY1976}
within a starting position. Unfortunately, in that case, the solution
depends heavily on a good estimation of the starting condition. In
order to detect the outlier and to obtain a strong starting conditions,
one needs to analyze the data. The general approach for the LPM to
detect an outlier was using Chi-test within a linear Kalman filter
(LKF) \cite{R.Pfeil2009} on raw data. Figure \ref{fig:LPM-raw-data}
shows that the offset ($O$) is $10^{6}$ times higher than the measurement
itself. Furthermore, the offset is changing from one measurement to
the next, due to the difference in oscillator clocks of reference
station and transponder \cite{R.Pfeil2009}.The elimination of the
offset allows us to see the relative range changing between the transponder
and base station. This can be done by subtracting one base station
from another at the same measurement. In figure \ref{fig:Relative-position}
the result of this transformation on real measurements can be seen.
The outliers are now visible with their typical characteristic of
reflected data. At this point the TOA equation is changing to the
TDOA. In \cite{Shin2002} it was proven that the error propagation
of the TDOA is equal to that of the TOA. This transformation can be
done for two purposes: first, one may obtain the relative range between
the base station and eliminate the offset; in this way we are able
to filter data and outliers; second, one can rearrange the TOA equation
to get the linear form of the transponder x, y, and z coordinates.

\subsection*{TOA to TDOA}

\subsection{Linearization}

Approaches such as the Taylor-series expansion \cite{FOY1976} or
other nonlinear solvers can be used to linearize the (\ref{eq:LPMeq})
equations. These methods require start information for the unknown
variables. Alternatively one may use a reference station to obtain
linear terms for the unknown position of the transponder. The offset
will still be a part of the equation, but, just like the coordinates
of the transponder, will be linear. This technique is also known as
linear least squares multilateration. The main LPM equation (\ref{eq:LPMeq})
can be simplified by adding the known reference transponder range
to the measurement term R (pseudo range). 
\begin{equation}
L_{i}=R_{i}+\left\Vert T-B_{i}\right\Vert 
\end{equation}

\begin{equation}
\left\Vert M-B_{j}\right\Vert ^{2}-\left\Vert M-B_{i}\right\Vert ^{2}=\left(L_{j}-O\right)^{2}-\left(L_{i}-O\right)^{2}
\end{equation}

The known quadratic terms of the transponder are eliminated, hence
the linear solution for the transponder position at the known base
station and reference station position is:
\[
\left(\overrightarrow{B_{i}}-\overrightarrow{B_{j}}\right)\cdotp\overrightarrow{M}-\left(L_{i}-L_{j}\right)\cdotp O=
\]

\begin{equation}
=\frac{1}{2}\left(\left(\left(\overrightarrow{B_{i}}\right)^{2}-\left(\overrightarrow{B_{j}}\right)^{2}\right)-\left(L_{i}^{2}-L_{j}^{2}\right)\right)
\end{equation}

with

\[
\begin{array}{cc}
\overrightarrow{M}=\left(\begin{array}{c}
x_{M}\\
y_{M}
\end{array}\right) & \overrightarrow{B}=\left(\begin{array}{c}
x_{B}\\
y_{B}
\end{array}\right)\end{array}
\]

The state-of-the-art Abatec LPM software uses a damped Newton iteration
for nonlinear regression \cite{Stelzer2004}, requiring some start
positions are required. The solution here presented, does not require
any start position or nonlinear solvers. The method\textquoteright s
result, however, depends on the condition of the coefficient matrix
(A). Should the pivots (diagonal elements of the coefficient matrix)
be close to zero the condition is bad. 

\subsection{Filtering}

\begin{figure}[t]
\begin{centering}
\includegraphics[scale=0.6]{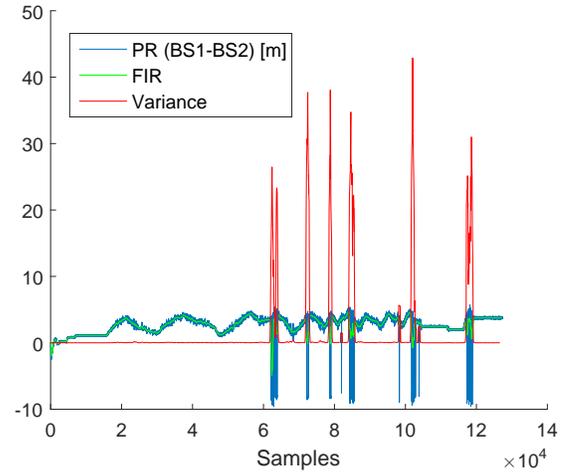}
\par\end{centering}
\caption{\label{fig:Reflections}Example of reflection inside real data. Red
curve represents variance, blue measurements and the green the filtered
data }
\end{figure}

\begin{figure}[t]
\begin{centering}
\includegraphics[scale=0.45]{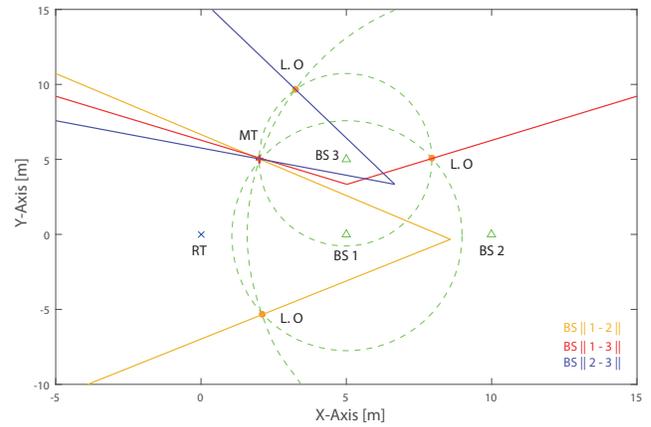}
\par\end{centering}
\caption{Difference between TOA and TDOA \label{fig:LPM-TDOA}. The triangles
are the simplified hyperboloids of TDOA and the green circles the
euclidean form of TOA }
\end{figure}

The transformation of the TOA to TDOA by subtracting one base station
from a reference station can also be used to filter data. If equation
\ref{eq:LPMeq} is not getting solved for T and squared before subtracted
from a reference station, the term will still be nonlinear for the
unknown position of the transponder, but the offset will be eliminated.\\
As mentioned above, every measurement has its offset, which is usually
$10^{6}$ times higher than the location itself. If the numerical
solver does not take this into account and the start conditions for
the offset are unfavorable, then the changing in x, y, and z coordinates
have almost no effect on the residuals. The higher the difference
between $||M-B||$ and offset $O$ the higher the deviation between
local optima and global optimum. Another problem could appear $R+||T\lyxmathsym{\textendash}B||>O$
in this case, the minimum of the numerical optimization becomes the
maximum. The most suitable starting value for the offset would be
the mean for every measurement, but we are still not able to set any
start condition for the coordinates of the transponder or to interpret
the measurements. In addition, the offset is changing from one measurement
to the next, hence it promises suitable to use the difference between
the base stations with respect to one reference station for the same
measurement for further calculations. This method is known as the
hyperbolic method \cite{Pourvoyeur2006} , due to the fact that the
pseudorange is not maintained from the perspective of the transponder,
if moving with a fixed radius on a circle around the basisstion, but
in order to maintain the same pseudorange movement has to remain on
a hyperbola. Changing the base station\textquoteright s instead the
of transponder\textquoteright s position would provide a hyperbolic
shape from the beginning. This shape is typical for TDOA, however,
now the multidimensional damped Newton is utilized to solve the minimization
of the equation \cite{Stelzer2004}. Data may be filtered before position
estimation, without the offset. The general approach for the LPM was
to use the extended Kalman Filter, (EKF)\cite{Pourvoyeur2006a} on
the data provided by the numerical solver. But even if the data of
the measurements has a gaussian noise, it does not necessarily mean
that numerical output has a gaussian distribution as well. The least
squares method to minimize the residuals $min\overset{n}{\underset{i=1}{\sum}}\frac{1}{2}(f(x,y,z)-y_{i})^{2}$is
effected by outliers, but is able to deal with gaussian noise. A bad
matrix condition or poorly chosen starting condition could lead to
a result, which does not have a gaussian distribution anymore and
therefore makes filtering the data before solving an advantage accompanied
by a solver that does not have to estimate the offset. The solver
converges faster and filtering can be one dimensional instead of three
dimensional. The difference between the base stations to one reference
station causes a dependency between the equations due to noise \cite{LukaszZwirello2012}.
Based on empirical measurements, the variance of the noise (figure
\ref{fig:,-Gaussian-Distribution:}) for every base station differs
from $0.003\,m{{}^2}$ to $0.0036\,m{{}^2}$ for an immobile transponder.
Thus one can assume that the variance is equal for every base station.
This is an important fact for iterative solving. There are several
filters applicable to filtering the data in question. A very fast
and simple filter is the moving average filter. This filter proves
helpful when the focus lies on the time domain instead of the frequency
domain. Its filter kernel (impulse response of the filter) does not
require a convolution with a signal, instead, processing can be reduced
to subtracting the oldest measurement and adding one of the latest
measurements. Due to the central limit theorem, running the filter
several times would lead to a similar result as using a convolution
with a gaussian kernel. In this way the over and under oscillations
of the frequency domain are reduced. The method is faster than the
convolution with a Gaussian kernel. In figure \ref{fig:Reflections}
the result of the moving average filter, which is a Finite Impulse
Response filter (FIR) illustrating the moving variance on real data
after the transformation. All outliers have a high variance with respect
to the other measurements. This information can be used not just to
detect the transponder, but also the source of the reflection. The
disadvantage of this method is a delay $(N-1)/2$ , which is increasing
with the number of samples $N$. As an alternative, one could use
the linear one dimensional Kalman filter instead. The difference between
TOA and TDOA is visible (fig.\ref{fig:LPM-TDOA}) Green circles represent
the probability of the transponder location. Every circle represents
one euclidean equation at a certain measurement. The correct location
fits all the equations, therefore the residuals are the smallest.
Smaller residuals are produced where just two equations fit, compared
to other positions, hence the found coordinates are the local optimum
(LO). The transformation results in simplified, hyperbola and triangles.
The position of the transponder can be found by two hyperbola, but
three hyperbola can be produced. The third can be represented by the
other two, this hyperbola has no further information compared to the
other two, but different local optima and another intersection angle.
Due to this fact, numerical solving may influence, just by transforming,
the probability to find a local optimum 
\begin{figure}[t]
\begin{centering}
\includegraphics[scale=0.6]{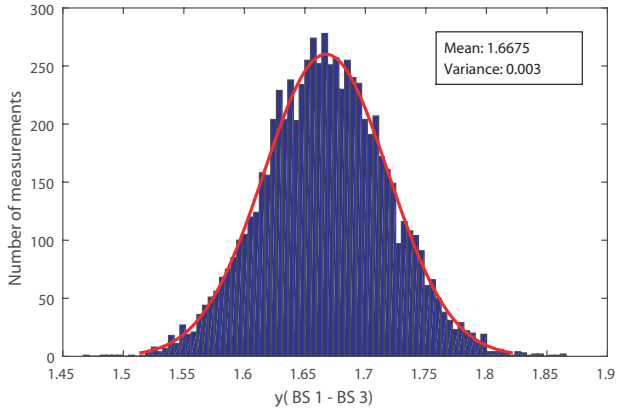}
\par\end{centering}
\caption{$BS_{1-3}$, Real data based gaussian distribution of the pseudo range
difference for a not moving transponder\label{fig:,-Gaussian-Distribution:}}
\end{figure}
\begin{figure}[t]
\centering{}\includegraphics[scale=0.5]{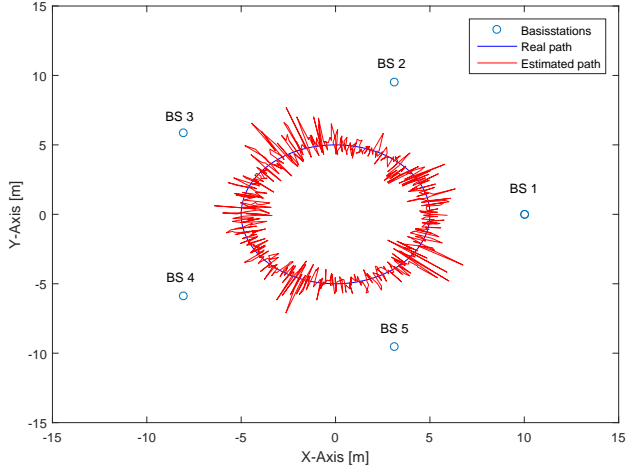}\caption{Synthetic data: Real path of transponder is located around the reference
station with a radius of 5~m. The base stations have a distance of
10~m from the center and same distance to each other. Measurement
data is corrputed by gaussian noise. \label{fig:Increase-in-noise}}
\end{figure}
 .

\subsection*{Combination of the linear equation with filtered data}

The linear equation has some advantages over the nonlinear solution.
This section presents a method of pre-filtering, which can be used
to correct the linear solution. Tests showed that the noise is ten
times higher than an immobile transponder, possibly due to the Doppler
effect and the micro movements of the carrier. Therefore, we increased
the random noise up to a maximum value of $0.1\,m$. The mean error
of the estimated path with respect to the real path is $0.506\,m$
(fig. \ref{fig:Increase-in-noise}). The measurement L1 cannot be
filtered as the time offset change with respect to time is $10^{6}$
times higher than the range change itself. Therefore, the offset is
eliminated by subtracting one base station measurement from the other
$(L_{i}-L_{j})$. In the next step this data is filtered over time.
At this point it does not matter what kind of filter is used, it is
only important that filtering takes place before position estimation
and that the filter uses the measurement difference $(L_{i}-L_{j})$
as an input. For the following calculations we only assume that for
every measurement we already have difference $(L_{i}-L_{j})$ the
filtered values $F(L_{i}-L_{j})$ the main aim is to use the filtered
values instead of the measurement differences between the base stations.
The $(L_{i}^{2}-L_{j}^{2})$ term is nonlinear but the filtered values
consist of the linear difference between the measurement ranges. One
solution to using the filtered values would be to make every base
station dependent on the same measurement error term $\alpha_{i}$.
Every measurement is corrupted by the measurement error \textgreek{a}i,
hence the real measurement can be written as $L=\tilde{L}+\alpha$.
The connection between the measurement errors \textgreek{a}i and \textgreek{a}j
can be found if the unfiltered measurement difference $(L_{i}+\alpha_{i})-(L_{j}+\alpha_{j})$
is subtracted from the filtered values $F(L_{i}-L_{j})$. \\
\begin{equation}
F_{ij}=\left(\left(\tilde{L_{i}}+\alpha_{i}\right)-\left(\tilde{L_{j}}+\alpha_{j}\right)\right)-F\left(L_{i}-L_{j}\right)
\end{equation}

\begin{equation}
\left(\tilde{L_{i}}-\tilde{L_{j}}\right)\thickapprox F\left(L_{i}-L_{j}\right)
\end{equation}

The assumption that the noise can be neglected after the filtering,
this leads to the term $F_{ij}$ being the difference between the
noises of both signals.\\
\begin{equation}
F_{ij}=\alpha_{i}-\alpha_{j}
\end{equation}
\begin{equation}
\alpha_{j}=-F_{ij}+\alpha_{i}\label{eq:replace}
\end{equation}

The measurement error is replaced by eq. \ref{eq:replace}.
\[
\left(\overrightarrow{B_{i}}-\overrightarrow{B_{j}}\right)\cdotp\overrightarrow{M}-\left(\tilde{L_{i}}-\tilde{L_{j}}+F_{ij}\right)\cdotp O=
\]

\[
=\frac{1}{2}\left(\overrightarrow{B_{i}}^{2}-\overrightarrow{B_{j}}^{2}-\tilde{L_{i}}^{2}-\tilde{L_{j}}^{2}-F_{ij}^{2}\right.+
\]

\[
\left.+2\cdotp\alpha_{k}\left(\tilde{L_{i}}-\tilde{L_{j}}+F_{ij}\right)\right)
\]

It can be observed that the time offset $O$, depends on the same
parameters as the measurement error
\[
\left(\overrightarrow{B_{i}}-\overrightarrow{B_{j}}\right)\cdotp\overrightarrow{M}-\left(\tilde{L_{i}}-\tilde{L_{j}}+F_{ij}\right)\cdotp\left(O+\alpha_{k}\right)=
\]
\[
=\frac{1}{2}\left(\overrightarrow{B_{i}}^{2}-\overrightarrow{B_{j}}^{2}-\tilde{L_{i}}^{2}-\tilde{L_{j}}^{2}-F_{ij}^{2}\right)
\]

With at least four base stations, the unknown coordinates of the transponder
can be estimated. With the filtered values the linear direct solution
provides better results, than with the unfiltered equation $Ax=b$.
This equation can be solved as:
\[
\left(\begin{array}{c}
x_{M}\\
y_{M}\\
O
\end{array}\right)=\left(A^{T}A\right)^{-1}A^{T}b
\]

\begin{figure}[H]
\centering{}\includegraphics[scale=0.45]{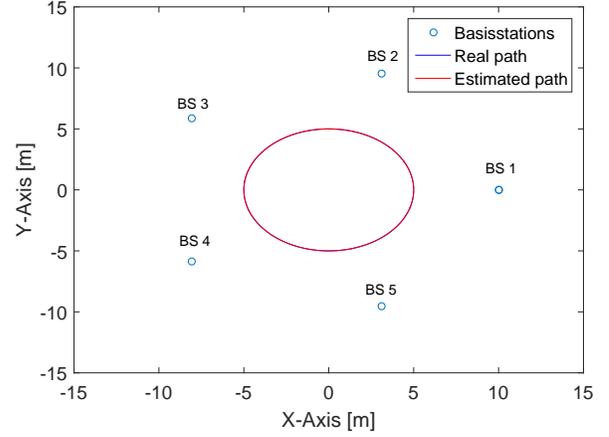}\caption{Noise correction\label{fig:Noise-correction}}
\end{figure}

Figure \ref{fig:Noise-correction} presents the results of the new
equation with noise. The mean error with a perfect filter is now below
$10^{-11}\,m$. The experiment was repeated with synthetic data and
a real moving average filter to investigate the behavior of outliers.
Simulating a second base reflection the sample was set from 3500 to
3600 an outlier of 10 meters. 
 
\begin{figure}[H]
\centering{}\includegraphics[scale=0.45]{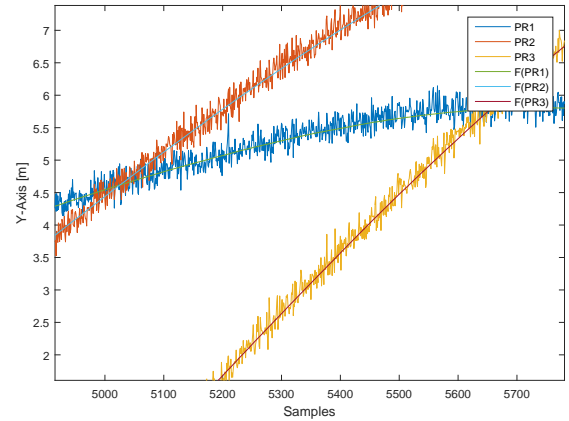}\caption{Synthetic data (PR) with gaussian noise and moving average filter
F(PR)\label{fig:Filter-zoom}}
\end{figure}
 
\begin{figure}[H]
\centering{}\includegraphics[scale=0.45]{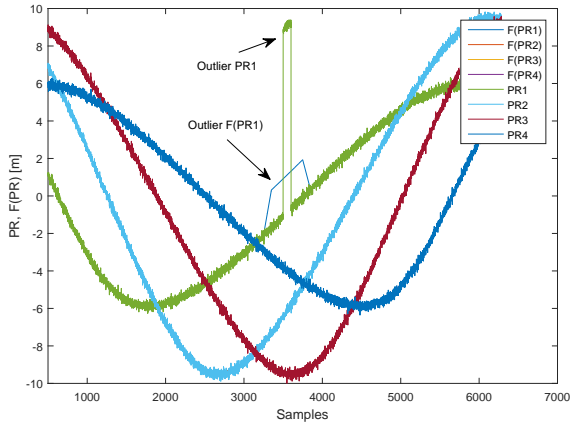}\caption{Filter error due to outlier: Such measurements are typical for signal
reflections \label{fig:Outlier}}
\end{figure}
\begin{figure}[H]
\centering{}\includegraphics[scale=0.45]{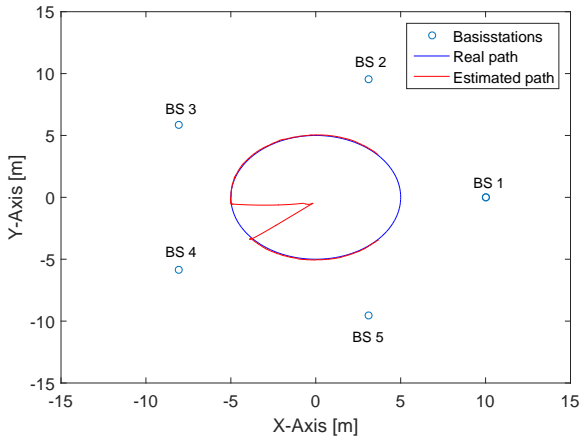}\caption{Outlier position estimation\label{fig:Outlier-position-estimation}}
\end{figure}
\begin{figure}[t]
\centering{}\includegraphics[scale=0.45]{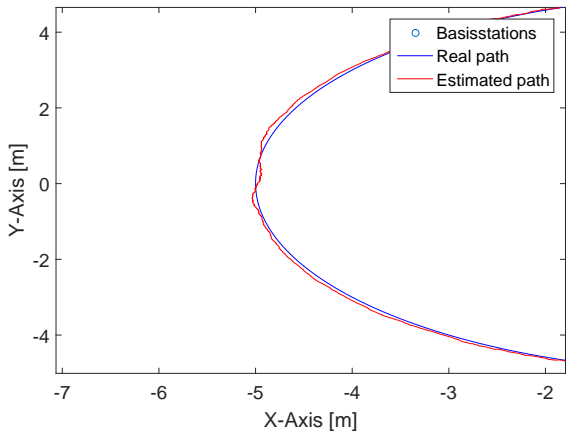}\caption{Result of the weighting on synthetic data with outlier\label{fig:Weighting}}
\end{figure}
\begin{figure}[H]
\centering{}\includegraphics[scale=0.45]{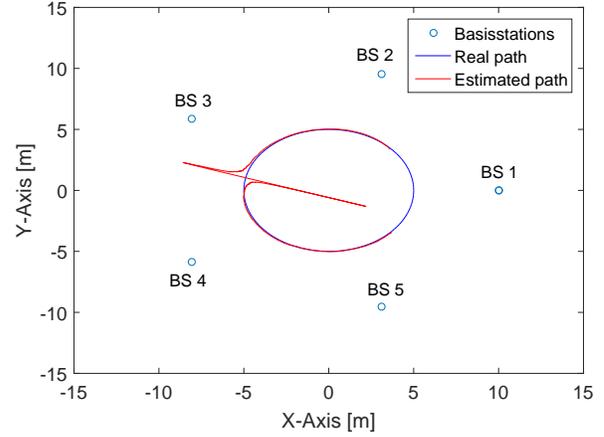}\caption{Result of weighting at unfavorable geometric condition\label{fig:Weighting-bad-circumstances}}
\end{figure}
Figure \ref{fig:Outlier} shows the results of the transformed signal
and the filtered one (figure \ref{fig:Filter-zoom}). The outlier
causes some disturbance in the filtered output, which leads to an
error in the position estimation (figure \ref{fig:Outlier-position-estimation}).
This filter error can be minimized by taking variance into account.
It is important to remember that the delay of the FIR filter to the
signal is $(N-1)/2$ and of the moving variance $(N-1)/2$ plus the
delay of the FIR, therefore we need some samples for the initialization
and have no position estimation at the beginning (fig. \ref{fig:Outlier-position-estimation}).
The provided variance is now used to obtain the weighting vector $w=1/var$.
\[
\left(\begin{array}{c}
w_{1}\\
\vdots\\
w_{n}
\end{array}\right)Ax=\left(\begin{array}{c}
w_{1}\\
\vdots\\
w_{n}
\end{array}\right)b
\]

\begin{figure}[t]
\centering{}\includegraphics[scale=0.45]{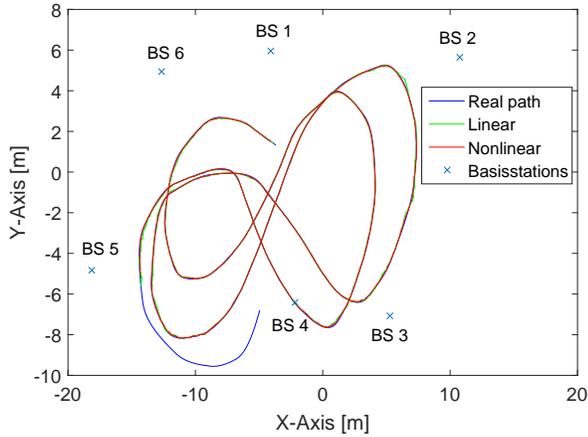}\caption{Linear and Nonlinear\label{fig:Linear-and-Nonlinear}}
\end{figure}

The result of the weighting is visible in figure \ref{fig:Weighting},
the measurement with the highest error now has the smallest weight
in the minimizing of the least squares problem. That being said, in
applying a working filter eliminates the need for weights. Applying
weighting at the position where the equation is sensitive to perturbations
like $(PR_{m}-PR_{n}+RT_{m}-RT_{n})\approx0$ or $B_{x},B_{y},B_{z}\approx0$
the result of the position estimation will be highly inaccurate, due
to corruption of the statistical meaning (fig. \ref{fig:Weighting-bad-circumstances}).
In conclusion, we need to compare the nonlinear and the linear solution.
Both methods have been transformed and filtered before the lateration.
As a nonlinear solver we have used a Levenberg Marquardt method (LVM)
\cite{More1977}. The starting condition for the solver was the result
of the previous fit beginning from the starting condition $x,y,z=0$
and fit parameter (table \ref{tab:New-linear-parameter}). Due to
the TOA to TDOA equation transformation, there is no need to fit the
offset, for every step LVM x, y, and z coordinates of the transponder
need to be estimated. 
\begin{table}[H]
\begin{centering}
\caption{New linear parameter\label{tab:New-linear-parameter} }
\par\end{centering}
\centering{}%
\begin{tabular}{|c|c|}
\hline 
Parameter & Value\tabularnewline
\hline 
\hline 
Scaled gradient & 0.000001\tabularnewline
\hline 
Relative function improvement & 0.00001\tabularnewline
\hline 
Scaled step & 0.001\tabularnewline
\hline 
Maximum iterations & 20\tabularnewline
\hline 
\end{tabular}\\
\end{table}
 Figure \ref{fig:Linear-and-Nonlinear} presents the results of both
methods. The linear solution has a slightly higher error rate compared
to the nonlinear solution in some positions. The linear calculation,
however, took $2596\,ms$ versus the nonlinear\textquoteright s $4677\,ms$
on a i7-4600u CPU 2.10GHz and 16 GB Ram. It follows that if prefiltering
would be a \textquoteleft perfect\textquoteright{} approach, both
solutions would have the same result as the real path. 
\\

\section{Conclusions}

The transformation from the TOA equation to the TDOA brings several
advantages for the LPM. Section 3 demonstrated that the transformation
leads to a linear equation, which can be solved without initialization
and numerical iterations. Admittedly, the linear solution as compared
to the nonlinear TDOA solution is more sensitive to perturbations
given specific geometric settings. Obtaining the condition of the
coefficient matrix, though, allows detection of these settings, and
permitting use of the faster linear solution for every other position.
The latter part of section 3 dealt with the TOA to TDOA transformation
with the aim to eliminate the offset and to filter the data before
the multilateration. Filtering the data before the position estimation,
because non-linear function estimation leads to violation of normal
distribution, made more sense, however. Furthermore, the transformation
allowed the detection of outliers, reducing the number of local optima
and decreasing the computational time due to the elimination of the
offset. In the end, the linear solution was expanded with the possibility
to use pre-filtered data and provide a solution at in half the time
that a non-linear solution can.

\bibliographystyle{plain}
\bibliography{bib_LMP}

\end{document}